%% file: ms.tex
\documentclass[10pt,sigconf,letterpaper]{acmart}

\newenvironment{itemize-s}%
{\begin{itemize}%
		\setlength{\itemsep}{0pt}%
		\setlength{\parskip}{0pt}}%
	{\end{itemize}}

\renewcommand\footnotetextcopyrightpermission[1]{} 
\usepackage[english]{babel}
\usepackage[utf8]{inputenc}
\usepackage[colorinlistoftodos]{todonotes}
\usepackage{epsfig,url}
\usepackage[T1]{fontenc} 
\usepackage{listing}
\usepackage{float}
\usepackage{color}
\usepackage{subcaption}
\usepackage{multirow}
\usepackage{balance}
\usepackage{cancel}
\usepackage{adjustbox}
\usepackage{epstopdf}
\usetikzlibrary{positioning,calc}
\usepackage{longtable}
\usepackage{siunitx}
\usepackage{enumerate}
\usepackage{enumitem}
\usepackage{rotating}
\usepackage{hhline}
\usepackage[printonlyused]{acronym}
\usepackage{tabularx}
\usepackage{booktabs}
\usepackage{soul}
\usepackage{xspace}

\usepackage{array}
\newcolumntype{x}[1]{>{\centering\arraybackslash\hspace{0pt}}p{#1}}

\input{sections/acronyms}

\newcommand{\UC}{{\sf devices-catalog}\xspace}

\newcommand{\ie}{{\it i.e.,}\xspace}

\newif\ifcomment
\commentfalse

\ifcomment
    \newcounter{AELNumberOfComments}
    \stepcounter{AELNumberOfComments}
    \newcommand{\ael}[1]{\textcolor{purple}{\small \bf [ael\#\arabic{AELNumberOfComments}\stepcounter{AELNumberOfComments}: #1]}}
    \newcounter{JinNumberOfComments}
    \stepcounter{JinNumberOfComments}
    \newcommand{\jin}[1]{\textcolor{orange}{\small \bf [jin\#\arabic{JinNumberOfComments}\stepcounter{JinNumberOfComments}: #1]}}
    \newcounter{AFNumberOfComments}
    \stepcounter{AFNumberOfComments}
    \newcommand{\af}[1]{\textcolor{red}{\small \bf [af\#\arabic{AFNumberOfComments}\stepcounter{AFNumberOfComments}: #1]}}
    \newcommand{\NOTE}[1]
    {
      {\footnotesize\it
        \begin{center}
          \begin{tabular}{|c|}
           \hline
            \parbox{0.85\columnwidth}{
              \medskip
              #1
              \medskip} \\
            \hline
          \end{tabular}
        \end{center}
        }
    }
\else
    \newcommand\dpnote[1]{}
    \newcommand\ael[1]{}
    \newcommand\jin[1]{}
    \newcommand\af[1]{}
    \newcommand\NOTE[1]{}
\fi

\renewcommand\footnotetextcopyrightpermission[1]{} 
\setcopyright{none}

\acmDOI{}

\acmISBN{}

\acmPrice{}

\begin{document}
\title{Where Things Roam \\ {\huge Uncovering Cellular IoT/M2M Connectivity } }

\author{Andra Lutu}
\affiliation{
	\institution{Telefonica Research}
}
\email{andra.lutu@telefonica.com}

\author{Byunjin Jun}
\affiliation{
	\institution{Northwestern University}
}
\email{byungjinjun2022@u.northwestern.edu}

\author{Alessandro Finamore}
\affiliation{
	\institution{Telefonica UK}
}
\email{alessandro.finamore1@telefonica.com}

\author{Fabian Bustamante}
\affiliation{
	\institution{Northwestern University}
}
\email{fabianb@cs.northwestern.edu}

\author{Diego Perino}
\affiliation{
	\institution{Telefonica Research}
}
\email{diego.perino@telefonica.com}

\renewcommand{\shortauthors}{X.et al.}

\begin{abstract}

Support for ``things'' roaming internationally has become critical for \ac{IoT} verticals, 
from connected cars to smart meters and wearables, and explains the commercial success of \ac{M2M} 
platforms. We analyze IoT verticals operating with connectivity via IoT SIMs, and present the first 
large-scale study of commercially deployed IoT SIMs for energy meters. We also present the first 
characterization of an operational M2M platform and the first analysis of the rather opaque associated 
ecosystem. 

For operators, the exponential growth of \ac{IoT} has meant increased stress on the infrastructure shared with traditional roaming traffic.
Our analysis quantifies the adoption of roaming by M2M platforms and the impact they have on the 
underlying visited \acp{MNO}. 
To manage the impact of massive deployments of device operating with an IoT SIM, 
operators must be able to distinguish between the latter and traditional inbound roamers. 
We build a comprehensive dataset capturing the device population of a large European \ac{MNO} over three 
weeks. With this, we propose and validate a classification approach that can allow operators to distinguish inbound 
roaming \ac{IoT} devices.

\end{abstract}

\maketitle

\section{Introduction} 
\input{sections/introduction}

\section{The Role of Roaming in IoT/M2M Connectivity} 
\input{sections/roaming_things}

\section{Dynamics of an M2M Platform}  
\input{sections/m2m_platform}

\section{View from an MNO} 
\input{sections/dataset}

\section{M2M Population Properties}
\input{sections/MNO_inbound_roamers}

\section{M2M Traffic Analysis}
\input{sections/m2m_traffic}

\section{The Case of Smart Meters}
\input{sections/smip}

\section{Discussion}
\input{sections/discussion}

\section{Conclusions}
\input{sections/conclusions}

\bibliographystyle{ACM-Reference-Format}
\bibliography{references}

\appendix
\section{Ethical considerations}
Both datasets used in this work are collected from operators and covered by NDAs prohibiting any re-sharing with 3rd parties even for research purposes. Raw data has been reviewed and validated by the operators with respect to GPDR compliance (e.g., no identifier can be associated to person), and all analysis performed report on aggregated metrics only.

\end{document}

%% file: sections/acronyms.tex
\acrodef{SGSN}{Serving GPRS Support Node}
\acrodef{GGSN}{Gateway GPRS Support Node}
\acrodef{CN}{Core Network}
\acrodef{RAN}{Radio Access Network}
\acrodef{xDR}{eXtended Detail Record}
\acrodef{TAC}{Type Allocation Code}
\acrodef{PoP}{Point of Presence}
\acrodef{IMEI}{International Mobile Equipment Identity}
\acrodef{EMSISDN}{Encrypted Mobile Station International Subscriber Directory Number}
\acrodef{MSISDN}{Mobile Station International Subscriber Directory Number}
\acrodef{MME}{Mobility Management Entity}
\acrodef{VoLTE}{Voice over \ac{LTE}}
\acrodef{MVNO}{Mobile Virtual Network Operators}
\acrodef{CDR}{Call Detail Record}
\acrodef{IMSI}{International Mobile Subscriber Identity}
\acrodef{RAT}{Radio Access Technology}
\acrodef{LPWA}{Low Power Wide Area}
\acrodef{VMNO}{Visited Mobile Network Operator}
\acrodef{HMNO}{Home Mobile Network Operator}
\acrodef{DCH}{Data Clearing House}
\acrodef{GSMA}{\ac{GSM} Association}
\acrodef{GSM}{Global System for Mobile communications}
\acrodef{TAP}{Transferred Account Procedure}
\acrodef{LTE-M}{\ac{LTE} Machine Type Communication}
\acrodef{MTC}{Machine Type Communications}
\acrodef{NB-IoT}{Narrow Band \ac{IoT}}
\acrodef{IOT}{Inter Operator Tariff}
\acrodef{IoT}{Internet of Things}
\acrodef{M2M}{Machine-to-Machine}
\acrodef{FNO}{Fixed Network Operator}
\acrodef{ISP}{Internet Service Providers} 
\acrodef{ASP}{Application Service Providers}
\acrodef{IPX}{IP Packet Exchange}
\acrodef{2G}{Second Generation}
\acrodef{3G}{Third Generation}
\acrodef{4G}{Fourth Generation}
\acrodef{5G}{Firth Generation}
\acrodef{ADB}{Android Debug Bridge}
\acrodef{ASCI}{Advertising Standards Council of India}
\acrodef{ASN}{Autonomous System Number}
\acrodef{CDN}{Content Delivery Network}
\acrodef{DL}{Downlink}
\acrodef{DNS}{Domain Name Service}
\acrodef{EaaS}{Experiment as a Service}
\acrodef{ECDF}{Empirical Cumulative Distribution Function}
\acrodef{HTTP}{Hyper Text Transfer Protocol}
\acrodef{ICMP}{Internet Control Message Protocol}
\acrodef{ISP}{Internet Service Provider}
\acrodef{IXP}{Internet Exchange Point}
\acrodef{LTE}{Long Term Evolution}
\acrodef{MBB}{Mobile Broadband}
\acrodef{MSC}{Message Sequence Chart}
\acrodef{E2E}{End-to-end}
\acrodef{QCI}{QoS Class Identifier}
\acrodef{NDT}{Network Diagnostic Tool}
\acrodef{QoE}{Quality of Experience}
\acrodef{QoS}{Quality of Service}
\acrodef{OS}{Operating System}
\acrodef{APN}{Access Point Name}
\acrodef{RMBT}{RTR Multithreaded Broadband Test}
\acrodef{RTT}{Round-Trip Time}
\acrodef{SIM}{Subscriber Identity Module}
\acrodef{SIMs}{Subscriber Identity Modules}
\acrodef{SLA}{Service-Level Agreement}
\acrodef{TCP}{Transmission Control Protocol}
\acrodef{UDP}{User Datagram Protocol}
\acrodef{UE}{User Equipment}
\acrodef{UL}{Uplink}
\acrodef{NRA}{National Regulatory Authority}
\acrodef{EC}{European Commission}
\acrodef{SGW}{Serving Gateway}
\acrodef{PGW}{Packet Data Network Gateway}
\acrodef{GTP}{GPRS Tunneling Protocol}
\acrodef{MNO}{Mobile Network Operator}
\acrodef{EU}{European Union}
\acrodef{HR}{home-routed roaming}
\acrodef{LBO}{local breakout}
\acrodef{IHBO}{IPX hub breakout}
\acrodef{VoIP}{Voice over IP}
\acrodef{IPG}{Inter-Packet Gap}
\acrodef{KS}{Kolmogorov-Smirnov}
\acrodef{FQDN}{Fully Qualified Domain Name}
\acrodef{MNC}{Mobile Network Code}
\acrodef{MCC}{Mobile Country Code}
\acrodef{MCCMNC}{\ac{MCC}-\ac{MNC}}
\acrodef{SIP}{Session Initiation Protocol}

%% file: sections/introduction.tex

%

The infrastructure established by Mobile Network Operators (\acp{MNO}) over 
the last 20 years for person-to-person communications is being leveraged to 
enable Internet of Things (\ac{IoT}) and Machine-to-Machine (\ac{M2M}) services.
In particular, support for ``things'' roaming internationally has become critical for \ac{IoT} 
verticals, from connected cars to logistic and wearables and explains the commercial 
success of \ac{M2M} platforms.

M2M platforms benefit from the extensive global network infrastructure that international carriers (e.g., incumbent tier-one operators such as Vodafone, Tata, Telefonica or Orange) have been shaping for the past decades, and created the so-called \ac{SIM} for things (or IoT SIM), which is a SIM provisioned by a single MNO, but operational anywhere in the world through roaming. 
This is attractive for IoT verticals, as using M2M platforms (i) can result in more stable 
connectivity/coverage, (ii) allow to avoid the cost of establishing technical and commercial relationships with operators in the countries they deploy, and (iii) application logic is handled in a centralized manner (all SIMs have a single home country) which can simplify management.
Moreover, the IoT SIM is compatible with any radio access technology advancement, such as \ac{LTE-M}, \ac{NB-IoT}, or even\ac{5G}, offering IoT verticals flexibility as they grow their deployments. 
This is particularly important given the predicted staggering growth of IoT deployment in the coming years.
However, this ecosystem remains largely unexplored in our community. 

In this paper we present the first characterization of the global 
footprint of an operational M2M platform and the first analysis of IoT SIM deployments in the wild, as part of this rather opaque roaming 
ecosystem. To do so, we take two different perspectives, as follows.

First, we present a characterization of the global footprint of an 
operational M2M platform, supporting IoT verticals in over 70 countries world-wide.
Using a 11-day long dataset and comprising 120k IoT SIMS, we expose both the ``centralization'' adopted by M2M platforms, 
as well as the breadth of their use across countries (Section~\ref{sec:m2m_platform}).

Second, we take the perspective of a visited \ac{MNO}, whose role (in this context) is to support these \ac{M2M} platforms and connect the IoT SIMs,
and we analyze the impact of roaming things on the \ac{MNO} infrastructure. 
We build a dataset that captures both real users and M2M/IoT devices of the large European \ac{MNO} over a period of 3 weeks (Section~\ref{sec:dataset}).
Our analysis quantifies the adoption of roaming by M2M platforms and the impact they have on the 
underlying visited \acp{MNO}. Out of 39.6M devices active across the 3 weeks, we find 26\% (10.1M) being M2M related, with 
75\% devices being international roamers (Section~\ref{sec:mno_population}).


The exponential growth of \ac{IoT} and current business models for international roaming 
is key also for \ac{MNO}s, as could lead to increased stress on the infrastructure shared with traditional roaming traffic.
Managing the growing stress of \ac{M2M} communication would not be a new problem for \ac{MNO}s 
if \ac{M2M} traffic showed similar characteristics to that of phone traffic (and brought comparable revenues). 
However, \ac{M2M} traffic exhibits significantly different features than phone traffic in a range of aspects from 
signaling, to uplink/downlink traffic volume ratios to diurnal patterns~\cite{shafiq2012first}.
In other words, though these devices occupy radio resources in \acp{MNO} networks and exploit the \acp{MNO} 
interconnections in the cellular ecosystem, they do not generate traffic that would allow \acp{MNO} to accrue 
revenue (Section~\ref{sec:m2m_traffic}). 

To manage the network and financial impact of \ac{M2M} traffic, operators must be able to distinguish 
between this and traditional inbound roaming traffic. This requires some ingenuity, and to support such task,
the GSM Association released a binding permanent reference document~\cite{lte-epc-roaming-guidelines}, recommending home networks and 
carriers to provide transparency of their outbound roaming \ac{M2M} traffic by sharing information on the dedicated 
APNs or dedicate IMSI ranges they use. In fact, if it is true that \ac{MNO}s should be able to identify their native devices,
\ie IoT devices that carry a MNO's SIM and connect to the MNO's infrastructure, without a common policy IoT devices
identification and classification is not an easy task. In this work we propose and validate a method based on both
device properties, traffic use, and APN strings.
We demonstrate this approach for the case of IoT \acp{SIM} deployed for energy smart meters (Section~\ref{sec:smip}). 



This paper makes the following contributions: 

\begin{itemize-s}
	\item \textit{We present the first characterization of mobile roaming support for M2M communication}.
	We describe how M2M platforms build on top of cellular infrastructure (\S~\ref{sec:roaming_m2m}), and 
	showcase the operation of a large M2M platform (\S~\ref{sec:m2m_platform}). We illustrate the sheer size 
	of these platforms with an analysis of the population of IoT SIMs activated/managed by it to support 
	IoT verticals over 4G networks.
		
	\item \textit{We show the impact of M2M roaming on (visited) \acp{MNO}}. We build a vast dataset to capture 
	the roaming status of all devices connected to a large European MNO for a period of 22 days (\S~\ref{sec:dataset}). 
	We introduce an approach for classifying devices into M2M, smartphones and feature phones. We present general 
	population characteristics, and show that the majority of IoT devices connecting to the MNO's network are roaming (\S~\ref{sec:mno_population}).
	
	\item \textit{We analyze IoT verticals operating with IoT SIMs, and present the first large-scale study of 
		commercially deployed IoT SIMs for energy meters}. We confirm that IoT SIMs' traffic patterns greatly differ 
	from those of smartphones (\S~\ref{sec:m2m_traffic}). We focus on smart energy meters, and present the largest 
	(3.2 million devices) measurement study of smart meters in real world deployments (\S~\ref{sec:smip}). 
	
\end{itemize-s}

%% file: sections/roaming_things.tex
\label{sec:roaming_m2m}

In this section, we expose how roaming supports cellular IoT/M2M communications. 
We close the section with a brief overview of related work in this space.

\subsection{Roaming Overview}

Roaming is one of the fundamental features of the cellular networks ecosystem. 
It enables clients of one \ac{MNO} to use the network of another \ac{MNO} 
when traveling outside the provider's area of coverage, nationally or 
internationally.

To support customers of a \ac{HMNO} roaming in the network of a \ac{VMNO} 
both networks must have a commercial agreement. With a technical solution in place, 
commercial roaming is then possible and MNOs' customers can use their respective 
partners' networks to extend coverage. MNOs generate roaming revenue by charging their 
roaming partners as a function of the data/voice/SMS the partner's users (inbound roamers) 
generate on the visited network. The roaming partners must each record the activity of 
roaming clients in a given \ac{VMNO}. Then, by exchanging and comparing these records, the 
\ac{VMNO} can claim revenue from the partner \ac{HMNO}.

In terms of business agreement solutions, the most popular option for \acp{MNO} is a
standard \emph{bilateral agreement} where the two parties involved define terms and conditions of their cooperation.
However, new bilateral roaming agreements for roaming are costly and are generally of lower value today. Even more, 
smaller and newer operators have great difficulty entering this market and extending their 
international coverage even for basic voice services. 

These challenges have motivated a new model that relies on \emph{roaming hubs}. In this model, operators connect to a hubbing 
solution provider to gain access to many roaming partners, externalizing the roaming interworking 
establishment to the roaming hub provider. 
Hubs are then interconnected to further expand potential operator relationships.  
The roaming hub solution does not preclude the existence of bilateral agreements between 
\acp{MNO}, and can be viewed as a complement to the bilateral roaming model.

\begin{figure}[t] 
	\centering
		\includegraphics[width=.98\columnwidth]{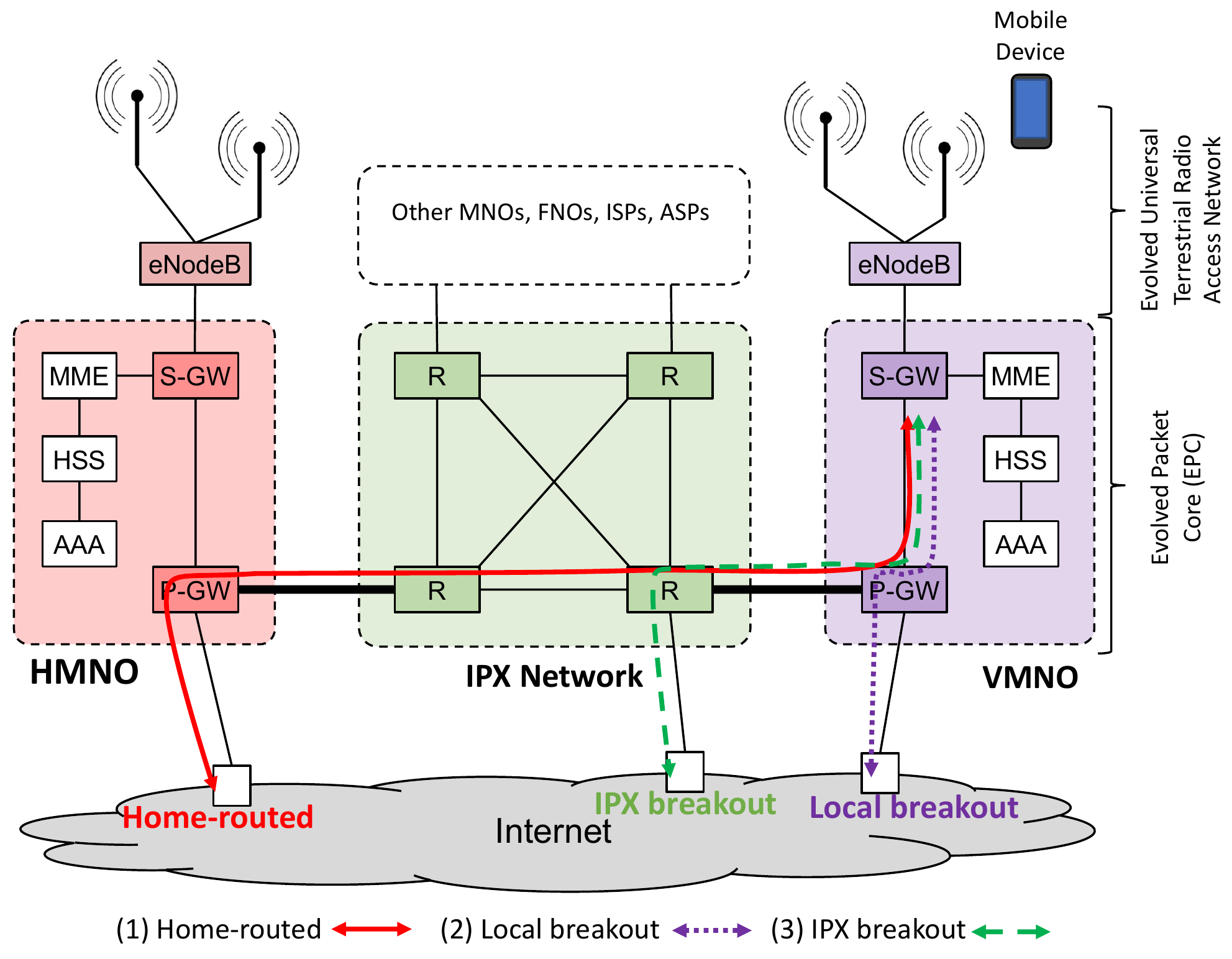}%
    \vspace{-3mm}
	\caption{\small Network configurations for roaming and interconnection of MNOs through a roaming hub (i.e., IPX Network).} 
	\vspace{-5mm}
	\label{fig:roaming}
\end{figure}

When a commercial agreement exists between two \acp{MNO}, there are multiple network 
configurations to enable roaming between the two networks. Figure~\ref{fig:roaming} presents 
a set of architecture configurations that can be used for roaming in a mobile network -- 
\ac{HR}, \ac{LBO} and \ac{IHBO}.
When a mobile node is at home (Fig.~\ref{fig:roaming}, left), the \textit{home user}'s traffic will take 
a short path inside the network to reach a suitable \ac{PGW} to the Internet. 
The traffic of a \textit{roaming user} (Fig.~\ref{fig:roaming}, right) is directed to an egress \ac{PGW} 
whose location depends on the roaming architecture.  
Previous work has found that the default roaming configuration currently used in majority of MNOs in 
Europe is the \ac{HR} roaming~\cite{mandalari2018experience}.

\subsection{Roaming for IoT/M2M}

All IoT device manufacturers need a global connectivity solution. 
This motivates them to evaluate communication providers who can ensure data connectivity across the globe, such as cellular connectivity providers. 
Roaming is thus an essential service for IoT verticals. Depending on the 
use case (e.g., automotive, logistic tracking, smart meters), roaming may be required occasionally or persistently/permanently. 
Different IoT verticals come with potentially different 
requirements -- while logistics services, for instance, may prioritize international roaming to track assets in flux, payment services depend on signal reliability, where terminals always connect, and select an alternative network in the event the first one fails. 

Adding to this complex scenario is the challenge of the relationship between \ac{VMNO}s and 
\ac{IoT} verticals. Whatever the constancy of the roaming requirement of IoT verticals, without any specific knowledge 
by the \ac{VMNO} of which inbound roamers on its network represent \ac{M2M} devices, the \ac{VMNO} may 
not be able to efficiently manage its relationship with the \ac{M2M} solutions and fully support these 
types of customers. The service specific levels of support required for roaming smart metering applications 
may differ from those for an e-book reader, or \ac{LPWA} devices. To support the applications efficiently, a 
\ac{VMNO} requires visibility of inbound roamers representing \ac{M2M} customers, dependent on what device or 
application is being used, so that it can assess the appropriate service impacts to support that \ac{M2M} 
roamer and manage the network efficiencies for M2M. Currently, transparency is provided by the \ac{M2M} 
\acp{APN}, \ac{IMSI} ranges (full or partial) and, for \ac{NB-IoT} (and other dedicated \ac{LPWA} platforms), 
the \ac{RAT}.

Given their strategic positioning as MNO interconnection providers and core function of ‘roaming hub’ 
(Fig.~\ref{fig:roaming}), international carriers have also morphed into providers for M2M 
businesses that need direct access to all the MNOs that connect to the roaming hub (e.g., Syniverse, 
BICS). They not only facilitate the technical relationship, but also the commercial relationships 
between MNOs, bringing the potential to serve \ac{IoT} players in every sphere and bridging 
the gap to seamless roaming. For example, BICS, one of the largest players in this space, 
interconnects with about 500 operators and carries about 25\% of worldwide roaming traffic, 
by its own estimates~\cite{bics2017}. These global 
carriers have an important role to establish reliable 
connectivity — so every vertical can access every place in the world through mobile connectivity, 
and manufacturers can produce a device in one part of the world that will connect to radio networks 
in another. 

Despite their growing importance, we have a limited understanding of the operational reality of M2M 
platforms dynamics and how \acp{MNO} actually support the IoT/M2M communications. A key contribution of 
this work is illuminating these aspects by analyzing two real-world datasets from an operational world-wide 
M2M platform and from an European MNO that hosts (i.e., as a VMNO) many devices whose connectivity is 
provided by the global M2M platform.


\begin{figure*}[!t]
    \centering
        \includegraphics[width=0.95\linewidth]{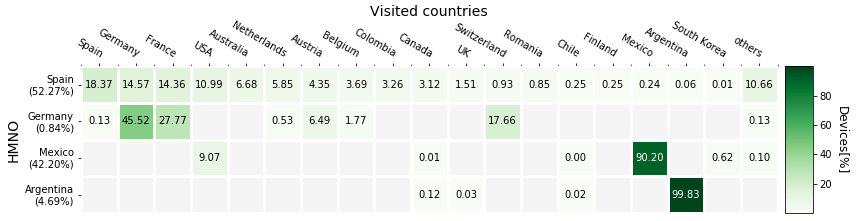}
    \vspace{-3mm}
    \caption{Percentage of M2M devices per visited country.}
    \vspace{-5mm}
	\label{fig:m2m_heatmap}
\end{figure*}



\subsection{Related Work}

Standardization bodies and different working group have been defining both network structure and services for M2M platforms~\cite{ETSIM2M,m2msurvey1,m2msurvey2,m2msurvey3}. Considering mobile networks, two opposite trends currently coexists, one pushing towards repurpose 2G/3G to serve M2M, and the other adopting 4G/5G~\cite{BIRAL20151,8030504,8095406}. Differently from this literature, we take a data driven approach focusing on the technologies we see deployed in live networks.

Furthermore, we core our analysis on roaming dynamics. Prior literature on cellular network traffic has focused on traditional, 
people-to-people communication or M2M communication within a single 
MNO~\cite{andrade:m2m,shafiq2012first,shafiq2013:m2m, kolamunna2018first}.  Vallina-Rodriguez et al.~\cite{vallina2015beyond}
analyzes roaming, primarily national roaming, using crowdsourced measurements. 
A recent study by Mandalari et al.~\cite{mandalari2018experience} presents an in-depth
characterization of international roaming in Europe, extending the work by Michelinakis et 
al.~\cite{michelinakis18infocom} limited to two operators. These past efforts have 
focused on roaming on traditional communication. There is however some literature regarding modeling M2M traffic~\cite{m2m-simulation1,m2m-simulation2,m2m-simulation3}, but it is intrinsically orthogonal with respect to our aim. Hence, to the best of our knowledge, this is \textit{the first study focused on roaming for M2M communication}.

%% file: sections/m2m_platform.tex
\label{sec:m2m_platform}

In this section, we focus on the operational system of a global M2M platform. This platform 
is built on top of an underlying international carrier and offers the service of global IoT \ac{SIM}.
The global IoT SIM is a SIM from a single (home) MNO that operates inside IoT devices world-wide through roaming. 
M2M platforms exploit roaming and the underlying carriers to give global connectivity to IoT providers, which
ship their devices internationally (from smart meters to wearables and cars) with pre-arranged cellular service.\footnote{In contrast
to an approach where IoT providers make local arrangements to obtain connectivity in \textit{each} country where their devices operate.}

The carrier that supports the M2M platform under consideration operates a large infrastructure world-wide, interconnecting directly with \acp{MNO} from 19 countries through 40 Points of Presence (PoP), with predominant presence in Europe and Latin America.
It further interconnects with other carriers to extend its footprint to the rest of the globe, and allow roaming on visited networks that are not directly interconnected to its PoPs.

We expose next the main characteristics of an operational M2M platform, focusing on 4G/LTE connectivity.
For this we use a dataset of passively collected signaling activity from IoT devices connected to networks world-wide through this platform (i.e., we do not capture traffic for 2G or 3G in the dataset).  


\subsection{M2M dataset}
The M2M dataset spans 11 days (November 19-29, 2018), and contains 14 million transactions generated by a population of over 120,000 4G-enabled IoT devices.
The monitoring probes capture control plane information, focusing specifically on the attach/detach procedures, as generated by devices connected to the VMNO radio network.
Given that few \acp{HMNO} issue the global IoT SIMs, the monitoring probes reside close to the infrastructure of the \acp{HMNO}. 
The dataset does not provide visibility into the data plane traffic, nor do we capture information on the specific IoT vertical served by the M2M platform.
Our goal here is to expose the reliance of the M2M platform on roaming and the carrier's roaming hub function to support IoT verticals on top of 4G/LTE networks. 
This provides visibility on the stress imposed by the dynamics of M2M devices on mobile networks, whose infrastructures offer the basic technological support for \ac{IoT}/\ac{M2M} services.

Each transaction in the signaling dataset reports an event generated by a IoT device attempting to connect the 4G radio network of an MNO, and the dataset represents a sampled view of world-wide M2M infrastructure traffic. 
More specifically, each transaction reports a unique device ID (a one-way hash), a timestamp, \ac{SIM} country code (~\ac{MCC}) and network code (\ac{MNC}), visited country code and mobile network code (\ac{VMNO} MCC-MNC), message type (either \emph{authentication}, \emph{update location} or \emph{cancel location}), and a message result (e.g., OK, RoamingNotAllowed, UnknownSubscription, etc.). 
In the remainder of the section, we evaluate the passive traces in this dataset to characterize the footprint of (4G) IoT device dynamics globally.



\subsection{Overall Dynamics}
\label{subsec:m2m_overview}

The records captured in the 11-days-long M2M dataset show that there are 4 main \acp{HMNO} that support M2M communications through the underlying infrastructure. 
To preserve anonymity, we denote them by their home countries, namely Spain (ES), Germany (DE), Mexico (MX) and Argentina (AR). 

 \begin{figure*}[th]
	\centering
	\begin{subfigure}{0.32\textwidth}
		\includegraphics[width=\linewidth]{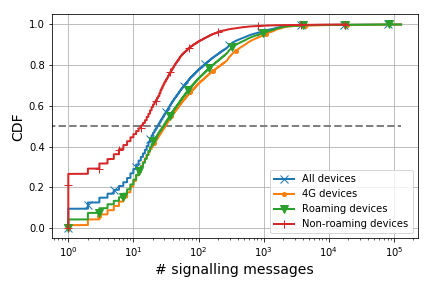}%
		\label{fig:m2m_records}%
	\end{subfigure}%
		\begin{subfigure}{0.32\textwidth}
		\includegraphics[width=\linewidth]{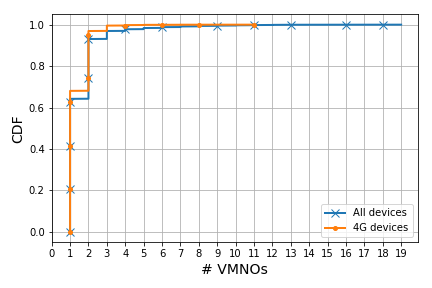}%
		\label{fig:m2m_vmnos}%
	\end{subfigure}%
		\begin{subfigure}{0.32\textwidth}
		\includegraphics[width=\linewidth]{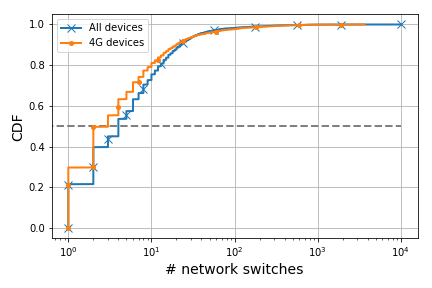}%
		\label{fig:m2m_network_switches}%
	\end{subfigure}%
	\caption{M2M Platform dynamics (left) distribution of total number of signaling records (center) VMNOs used; (right) inter-VMNO switches.}
	\label{fig:m2m_platform_dynamics}
\end{figure*}

Figure~\ref{fig:m2m_heatmap} presents the overall distribution of IoT devices on each of these \acp{HMNO}.
Each column shows one of the \acp{HMNO} and each row corresponds to the different visited countries where the IoT devices operate. 
We breakdown countries having at least 0.1\% of devices, and we group the rest into a single class ``Other''.
We normalize cell values by row, while the y-axis labels report the share of devices for each \ac{HMNO}.
We find that two \acp{HMNO} support majority of the M2M communications. 
In particular, the MNO from Spain provides the \ac{SIM} cards for 52.3\% of all the IoT devices in our dataset. 
Overall, during the entire period of analysis, the devices enabled by the M2M platform with \ac{SIM}s of Spain were active in 77 different countries, connecting to over 127 \acp{VMNO} through the M2M platform.
We note that the Spanish MNO is active in a region where free roaming has been promoted intensively through regulation~\cite{ec-roaming-charges}. 

The second most important \ac{HMNO} supporting the operations of the M2M platform is Mexico, with 42.2\% of all devices operating with a \ac{SIM} card belonging to this MNO. 
These IoT devices spread in 7 countries and connect to 10 \acp{VMNO} overall.
Note, however, that the large majority (90\%) operate in their home country and are not roaming. 
We conjecture this is due to the local restrictions on roaming in countries in Latin America. 
Argentina, much in a similar manner to Mexico, has 4.7\% of devices (with 6 visited networks) and almost all of its traffic are not roaming.

The fourth \ac{HMNO} we identify, the German MNO, supports a relatively small number of devices (around 1,000), but the number of visited networks is large (18 VMNOs).
This might be explained by the requirements of the specific IoT vertical.
For example, connected cars have high mobility requirements that would explain the need for seamless coverage, thus generating numerous signaling procedures from the devices and requiring alternative connectivity from multiple networks~\cite{andrade2017connected}.

Given that the Spanish MNO supports a large portion of IoT devices in our dataset, we continue our analysis on the dynamics of the M2M platform by capturing only IoT SIMs this \ac{HMNO} provides, which is either local (non-roaming) or global (roaming).\jin{Previously we indicates only "global", and thus it could confuse some readers. Now we explicitly show that both are visible in our dataset.}
For the Spanish network, roaming extends coverage over 76 countries and also generates large amounts of signaling traffic (81.8\% of all signaling traffic in our dataset comes from ES-powered IoT devices).
We verify that 92\% of these messages are triggered while devices are roaming. 
Conversely, only 8\% of the signaling traffic we capture from these devices occurs when they attach to the \ac{HMNO}, even though the fraction of non-roaming device is relatively high (18\%).
This suggests that IoT devices active in their native home country are potentially less mobile than the roaming ones and are connected over longer periods of time. 

For the roaming devices supported by the Spanish MNO (82\%), we find that 75\% of the signaling traffic comes from 62\% of devices. 
This covers operations over only 5 visited countries and 10 visited MNOs. 
The geographical distances between the HMNO and the VMNO are not always small (e.g., Spain to Australia), pointing to potential serious performance penalties in the case of \ac{HR} roaming~\cite{mandalari2018experience}. 
In this case, the M2M platform uses different roaming configurations in order to optimize the performance of IoT devices roaming in very far destinations. This analysis is, however, outside the scope of this work. 



\subsection{Device-level Dynamics}
\label{subsec:m2m_client}

We now focus our analysis on the device-level signaling traffic patterns of IoT devices connecting with a global IoT SIM for the Spanish provider. 
Specifically, we look at the frequency of three procedures we monitor (Update Location, Authentication and Cancel Location). 
Each record has a status message associated, describing the outcome of the procedure (i.e., OK, Feature Unsupported, Roaming Not Allowed or Unknown Subscription). 
We find that in this IoT device population, 40\% of devices trigger failed signaling procedures against the 4G/LTE networks.   
For the rest of 60\% IoT devices connecting through the Spanish MNO, we register at least one successful procedure in our dataset.
This is a non-negligible number of IoT devices generating traffic through the 4G signaling infrastructure by attempting (and failing) to use 4G connections. 
 


We further investigate the amount of signaling traffic per roaming IoT device, the distribution of number of \acp{VMNO} used and the frequency of inter-VMNO switches (see Figure~\ref{fig:m2m_platform_dynamics}).
First, we note that the distribution of the number of signaling records per device has a long tail, showing the wide range of signaling patterns the M2M dataset captures 
(Fig.~\ref{fig:m2m_platform_dynamics}-left).
We show this distribution for all IoT devices, and for devices successfully connected to the 4G network (4G devices), roaming devices and non-roaming (native) devices. 
From the distribution on all devices, the average load is of 267 signaling records overall, with 97\% of devices triggering less than 2,000 signaling procedures over the 11 days period, and a very small fraction of IoT devices flooding the signaling network with as many as 130,000 messages. 
We note the difference between roaming and native devices, with the former generating 10 more procedures than the latter in median.


Figure~\ref{fig:m2m_platform_dynamics}-center
shows the number of VMNO the roaming IoT devices use over the observation window. 
We find that 65\% of roaming IoT devices use only one VMNO, while more than 25\% roaming IoT devices switch between two \acp{VMNO}. 
Only 5\% of roaming devices require coverage from more than three VMNOs. 
Interestingly, for some of the IoT devices with only failed signaling procedures, we find that the maximum number of attempted VMNOs is as high as 19 mobile networks. 
This shows high international mobility requirements and the need for reliable seamless coverage, which are indeed difficult to guarantee with only 4G/LTE connectivity in some regions. 
We conjecture that these devices fall-back on 2G/3G coverage (which our dataset does not capture).


Beyond the number of VMNOs, we are also interested in the frequency of switches of a global IoT SIM between \acp{VMNO}. 
For those IoT devices with at least two VMNOs (35\% of IoT devices), we examine the number of inter-MNO switches. 
Figure~\ref{fig:m2m_platform_dynamics}-right shows a mixed result.
For approximately 50\% of IoT devices, we register a maximum of two VMNOs switches during the total 11 days.
However, for 20\% of devices, the inter-VMNO switches happen at least once a day.
Approximately 3\% devices present high frequency in switching between VMNOs, namely from 100 times to 3,000 times during the period we monitor. 
Again, we do not have visibility into the IoT vertical using these devices, but their high mobility and their requirements for reliable coverage is clear.




%% file: sections/dataset.tex
\label{sec:dataset}

\begin{figure}[!t]
	\centering
	\includegraphics[width=\linewidth]{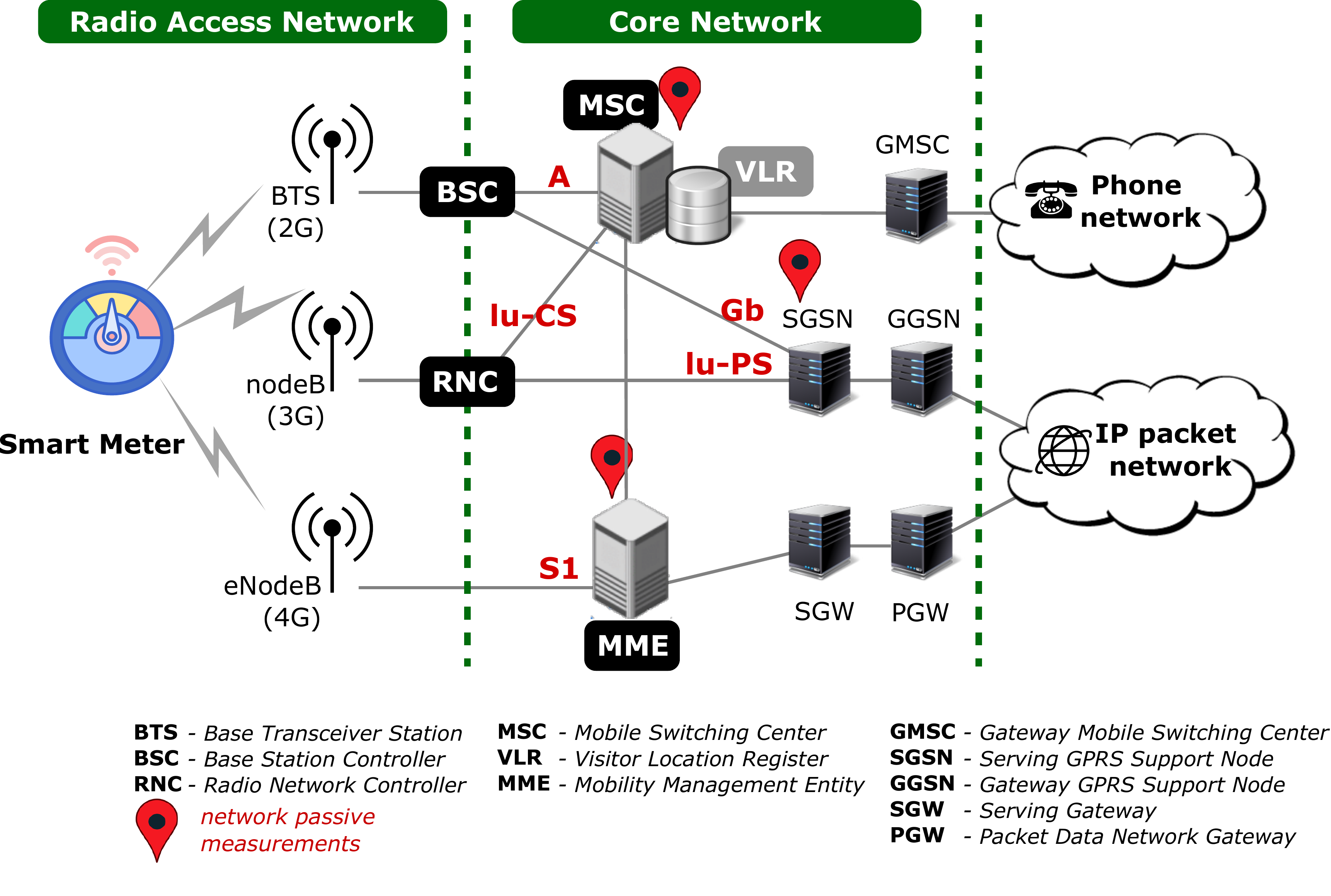}%
	\caption{High-level architecture of the measurement infrastructure integrated in the cellular network. }
	\label{fig:mno_architecture}
\end{figure}

%

Our analysis so far offered a global-scale view of the breadth and operations of an M2M platform (\S~\ref{sec:m2m_platform}). 
In this section, we focus our analysis on IoT/M2M communication dynamics from the point of view of one operational cellular network in the UK, which supports a large number of IoT devices as inbound roamers from the M2M platform.

\subsection{MNO dataset}


The cellular network we study supports 2G, 3G and 4G mobile communication technologies.  
In Figure~\ref{fig:mno_architecture}, we illustrate a high-level schema of the MNO architecture.
Such a network can be simplified to consist of three main domains: (i) the cellular device (in our case, the smart meter), (ii) the \ac{RAN} and (iii) the \ac{CN}.
Our passive measurement approach relies on commercial solutions integrated within the MNO's infrastructure.
The red pins in Fig.~\ref{fig:mno_architecture} mark the network elements that we monitor, namely the \ac{MME}, the \ac{MSC} and the \ac{SGSN}.
We collect control plane data on the total population of devices connected to the MNO's radio network. 
This includes both native devices (operating with a SIM card provisioned by the UK MNO) or inbound roaming devices (operating with a SIM card provisioned by a foreign MNO, from outside the UK, such as the global IoT SIM provided by the M2M platform). 

The dataset we use provides a comprehensive view on the entire population 
of mobile devices connected to the MNO's network over a period of 22 days in April 2019 
(from April 5th to April 26th 2019).
These include both smartphones and feature phones, as well as devices deployed for IoT verticals (e.g., smart meters). 
This population also integrates users with different roaming status, including MNO's \emph{native} users (active either in the home country or abroad), the users of MVNOs that operate on top of the MNO's infrastructure, and foreign users that belong to other MNOs (national or international), but that use the radio network of the MNO under analysis.

We start introducing the raw data collected by the operator, and then we discuss
how we label devices based on their roaming category, and how we identify IoT devices.


\noindent{\bf Radio interfaces.}
We process logs reporting on activities on IuCS, IuPS, A, and Gb radio interfaces.
Those carry events generated by the devices connecting to the radio sectors,
and requesting resources for either data or voice communications.
Each event carries the anonymized user ID, \ac{SIM} \ac{MCC} and \ac{MNC}, \ac{TAC}\footnote{The first 8 digits of the device IMEI, 
which are statically allocated to device vendors.}, the sector ID handling
the communication, timestamp, event type, event result code. 
Such events are captured for all connected devices, except
for outbound roamers (in this case, radio signaling for outbound roamers is carried over the visited country
network only).

\noindent{\bf Service usage.}
We use \acp{CDR} and \acp{xDR} to 
provide aggregate service usage for calls and data. 
Each record reports the anonymized user ID, MCC and MNC codes for both device SIM and visited country, 
timestamp, duration, and bytes consumed. Data records also report \ac{APN} strings, which 
usually encode information about the specific service/business they relate to.
Notice that differently from radio logs, CDRs/xDRs contain traffic also for outbound roamers. 
These are usually used by the roaming partners to trigger the process of revenue retrieval from roaming (see \S~\ref{sec:roaming_m2m}).

\noindent{\bf Device properties.}
We also consider a commercial database provided by GSMA. 
This catalog maps the device \ac{TAC} to a set of device properties such as 
device manufacturer, brand and model name, operating system, and radio bands supported.

\noindent{\bf Daily \UC.}
We combine the three data sources to create a daily list of active devices and associated
properties and traffic characteristics. We refer to this daily aggregate view as \UC.
Each record in the generated catalog reports a device ID, total number of events, calls, bytes seen, \ac{SIM} MCC/MNC, list of
visited MCC-MNC, list of \ac{APN} strings, device manufacturer, device model, device \ac{OS}.
We further summarize the radio activity into \emph{radio-flags}, a series of three 1-bit flags which
are set to 1 if the device has successfully communicated with 2G, 3G, 4G sectors respectively on radio interfaces.
Finally, we compute mobility metrics for each device.
Specifically, from radio logs, we compute the time spent on each individual sector to which a device connected.
Then, we use it to compute a weighted centroid and gyration, using sector coordinates provided by 
the MNO sectors catalog.


\subsection{Roaming Labels} 
\label{sec:roaming_labels}

To capture the roaming status of the MNO's population, we label each device as either 
\emph{native}, \emph{inbound} roamer, or \emph{outbound} roamer.
A device is native if it carries an MNO's \ac{SIM} and connects to that same MNO's radio network.
When such devices connect to a different operator network (either within the same country, or
when traveling outside the country) they become outbound roamers (national or international, respectively).
Conversely, an inbound roamer is a device operating with a \ac{SIM} card not belonging to the MNO whose radio network is actually using.

To capture these variants, we tag each record in the \UC with a \emph{roaming label}
\texttt{<X:Y>}, where \texttt{X} relates to the device SIM, and \texttt{Y} to the visited network.
Specifically, given a \ac{SIM} card, we assign to \texttt{X} four possible values:
{\bf H} ({\it home}, the SIM belongs to the MNO we analyse), {\bf V} ({\it virtual}, 
the SIM belongs to an MVNOs enabled by the MNO we analyse), {\bf N} ({\it national}, the SIM 
belongs to another MNO in the same country as the current MNO), {\bf I} ({\it international}, 
the SIM belongs to an MNO in a country different than the one of the MNO under study).
Instead, we assign to \texttt{Y} only two values: {\bf H} ({\it home}, 
the SIM is attached to the current MNO), {\bf A} ({\it abroad}, the SIM is attached to a 
foreign MNO outside the country of the MNO under study).

Overall, we define 6 different roaming labels. For example, the {\tt H:H} label reflects a device 
that uses the MNO's SIM card and is attached to the MNO's network (i.e., native user), while the {\tt I:H} 
label refers to devices connected to the MNO under study but with a SIM of operators of different countries than the MNO one (i.e., inbound roamers). 
Using these labels, we further breakdown devices per roaming status. 
As expected, we find that the majority of devices are native, i.e., either MNO (about 48\% per-day) or MVNO (about 33\% per-day) devices
connected to their home MNO.
However, we find that the third largest population are international inbound roamers (about 18\% per-day). 
The shares of devices of the roaming labels are stable across the 22 days we verify.

\subsection{M2M Device Classification}
\label{sec:device_classification}

As previously mentioned, the \UC includes all devices connected to the MNO network.
This encompasses devices used by people as their main personal device (e.g., smartphones, feature phones),
as well as devices \ac{IoT} verticals use to support their applications (e.g., car manufacturers, energy companies).
Based on our analysis from the point of view of the M2M platform (\S~\ref{sec:m2m_platform}), we aim to show that IoT devices are usually roaming internationally. 
To do so we require to split the devices into three classes: \emph{smart} (for smartphones), \emph{feat} (for feature phones), and \emph{m2m} (for IoT/M2M devices). 
Prior work~\cite{shafiq2012first} demonstrated that using device properties one can perform a classification, especially to spot M2M devices, at the cost of some manual verification. The GSMA database already offers a device classification label, but devices other than smartphones are mostly marked as ``modem'' or ``module'' which might not necessarily imply an M2M/IoT application.
Furthermore, across the 22 days we observe 2,436 device vendors, and 24,991 device models across the whole population (i.e., a manual classification as operated in~\cite{shafiq2012first} is not feasible).

One possible approach to reduce the classification complexity is to focus on ``big players'' only.
For instance, Gemalto, Telit, and Sierra Wireless, are among the top device vendors with a combined 75\% of all inroaming devices in the dataset.
Gemalto is a company with a broad portfolio of solutions for M2M/IoT communications;
Telit is a global leader in IoT enablement and well-known as a wireless M2M modules vendor;
Sierra Wireless is a multinational wireless communications equipment designer active in the M2M domain.
Similar considerations can be made to identify smartphones and feature phones, but we argue that this is a na\"ive approach as still requires to investigate a large number of devices to validate the classification.

APNs string can be a significant aid for strengthening the confidence of the classification as they hint the vertical used by a device.
For instance \emph{smhp.centricaplc.com.mnc004.mcc204.gprs} hints to Centrica\footnote{\url{https://www.centrica.com/about-us/what-we-do/our-strategy}}, a company working in the energy vertical, i.e., the devices using such APN are possibly smart meters. Notice also the MCCMNC revealing the home country and operator (20404 = Vodafone Netherlands).

We find a total of 4,603 APN strings in the dataset. However, ranking the APNs by number of devices using it, we identified 26 ``keywords'' in the APN string which we mapped to M2M/IoT verticals using information found online (e.g., \emph{scania} - automotive company, \emph{rwe} - energy company, \emph{intelligent.m2m} - global IoT SIM provider). Using these 26 keywords we obtained 1719 APNs, while the other are either generic string related to mobile operators (2,178 string likely related to consumer services), or other IoT services we could clearly identify.

Finally, we classify the devices combining both APNs and device properties. We start marking as \emph{m2m} all devices using the validated APNs. Then, we extend the \emph{m2m} class to all devices having the same properties of the devices using the validated APNs. For \emph{smart} and \emph{feat} we still use a set of APN strings, but we take advantage of 2 labels properties defined in the GSMA database. Specifically, we classify a device as \emph{smart} if declared to be using a major smartphone OS (android, iOS, blackberry, windows mobile) and use a consumer APN (e.g., a string contain keywords such as \texttt{payandgo}). Instead, we classify a device as \emph{feat} if the GSMA database declare it to be a feature phone or uses a consumer APN.

Out of the 39.6M devices active across the 22 days, we find 24.4M (62\%) \emph{smart}, 3.1M (8\%) \emph{feat}, and 10.1M (26\%) \emph{m2m}. We label the remaining 2M (4\%) as \emph{m2m-maybe} as the device properties suggest they are neither smartphones nor feature phones, but we don't have APNs for them, i.e., those devices only use voice services (the APN is provided only when the device connect for data services). This does not preclude them from possibly being M2M related, but based on the information available we are not able to provide a final classification. Hence, we do not consider those devices for the remainder of the analysis.

Differently from~\cite{shafiq2012first}, using APNs is useful to increase the classification confidence, and reduce the amount of manual investigations. However, when used in isolation, APNs are not enough as we find about 21\% of the devices in the dataset not having any APN. This justifies our multi-steps classification process (keywords$\rightarrow$APNs$\rightarrow$device properties).

\subsection{Smart Meters Dataset}
%

In this section, we describe the dataset we build specifically for smart meter devices connected to the radio network of the MNO we monitor. 
We use this dataset to focus our analysis of the roaming of things on a specific vertical (i.e., energy) and compare with the analysis of the general population of devices or other vertical (i.e., connected cars). 
Smart Grid applications have received increasing attention in the past years, with regulation pushing for mandatory deployment of metering devices in consumer premises. 
Specifically, the UK Government is committed to ensure that every home and small business in the country is offered a smart meter by 2020-2021, with more than 12 million device already deployed at the end of 2018\cite{ofgem2018smip} as part of the Smart Metering Implementation Programme (SMIP).

The mobile operator we study provides connectivity for a set of smart meters in the UK under the SMIP framework (\ie SMIP native devices). 
Based on private communications, we learned that the MNO uses a dedicate \ac{IMSI} range for the \acp{SIM} installed in smart meters.  
Moreover, the operator has dedicated resources for the \ac{GGSN} for these \acp{SIM}. 
The rationale of this choice is to control the impact of such devices on the native users as well as better control performance of the smart meter network.
We further denote this dataset as \textit{SMIP native}.

We further investigate whether there are other devices connecting to the MNO's radio network that are smart meters. 
Specifically, we aim to identify the use of Global IoT SIMs to connect smart meter devices. 

\begin{figure*}[thb]
	\centering
	\begin{subfigure}{0.85\textwidth}
		\includegraphics[width=\linewidth]{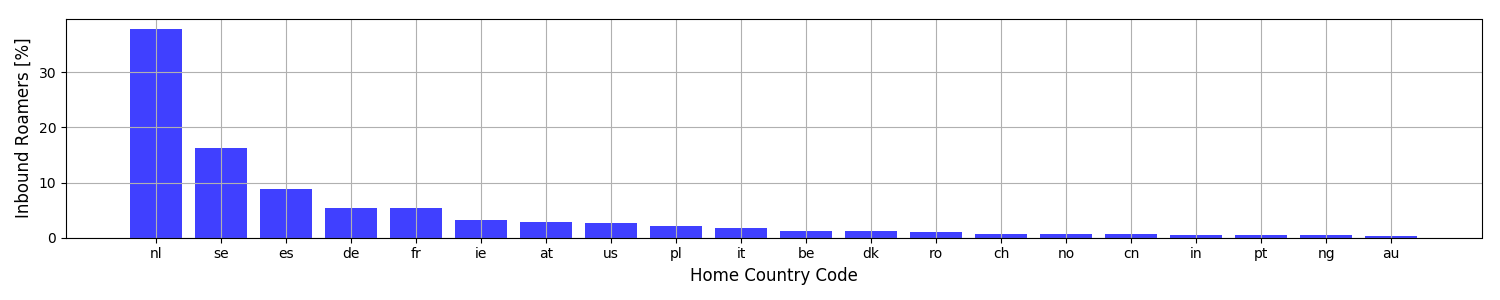}%
		\label{fig:distro_homeCountry}%
	\end{subfigure}%

	\begin{subfigure}{0.85\textwidth}
		\includegraphics[width=\linewidth]{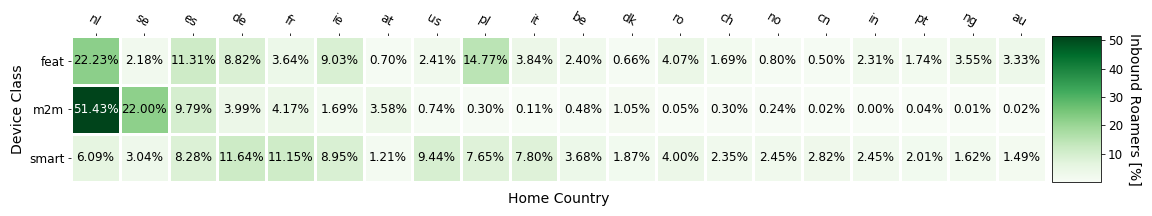}%
		\label{fig:home_country_device_class}%
	\end{subfigure}%
	\caption{Home country of inbound roaming devices (top) overall share; (bottom) device class breakdown.}
	\label{fig:homeCountry}
\end{figure*}

To identify these devices, we rely on the classification we explained above. 
Specifically, in the \ac{APN} strings of these inbound roamer devices we are able to identify patterns that confirm the use of these devices as smart meters by energy companies in the UK.
We are able to identify different patterns in the Network Identifier part of the \ac{APN} string that relate to five of the largest energy companies in the UK, namely Elster, RWE, Centrica PLT, General Electric and BGLOBAL Services.  
Using these, we are able to separate the inbound roaming devices that are likely smart meters. 
Surprisingly, all the \acp{SIMs} we identify are provisioned by the same cellular operator in the Netherlands. 
To further validate our inference, we use the \ac{GSMA} \ac{TAC} data catalog to identify the manufacturers of these devices. 
We find that these devices map to only two manufacturers mainly specialized in M2M modules, namely Gemalto and Telit. 
We further denote this dataset as \textit{SMIP roaming}.

%% file: sections/MNO_inbound_roamers.tex
\label{sec:mno_population}

 \begin{figure}[!t]
	\centering
	\begin{subfigure}[]{0.2\textwidth}
		\includegraphics[width=0.95\linewidth]{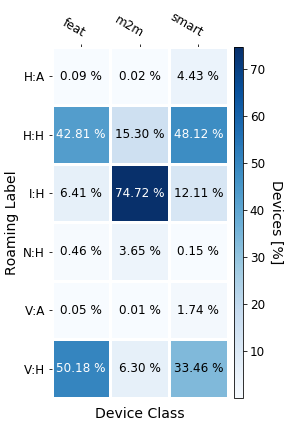} 
		\label{fig:device_class_aggregation}
	\end{subfigure}
	\begin{subfigure}[]{0.2\textwidth}
		\includegraphics[width=0.95\linewidth]{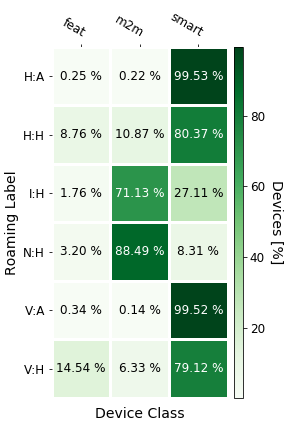} 
		\label{fig:roaming_cat_aggregation}
	\end{subfigure}	 
	\caption{Devices breakdown (left) Device class -vs- Roaming label; (right) Roaming label -vs- Device class.  
	\label{fig:device_class}}
\end{figure}


With the processed dataset, in this section we investigate the home country of the devices, if they are constantly
connected to the (visited) MNO network, and if they are stationary or moving.
To better highlight those properties, we contrast M2M devices against smartphones and feature phones.

\subsection{Device Class and Roaming Label}
Figure~\ref{fig:device_class} shows heatmaps of the distribution of devices per roaming label and per device class, 
normalized by device class 
(Fig.~\ref{fig:device_class}-left)
and by roaming category 
(Fig.~\ref{fig:device_class}-right).
Considering inbound roamers (\texttt{I:H}), 71.1\% are M2M device, while 27.1\% are smartphones (right heatmap). 
This further supports the popularity of supporting IoT verticals on top of the roaming infrastructure of cellular 
providers (\S~\ref{sec:m2m_platform}).  

Considering the device classes (left heatmap), 74.7\% of M2M are inbound roaming, while the rest are either native (\texttt{H:H}) or related to the MVNO (\texttt{V:H}). Instead, for smartphones and feature phones the trend is almost reversed: only 12.1\% and 6.4\% respectively are inbound roaming, while those device classes are either native of MVNO related.


\begin{figure}[!t]
	\centering
	\begin{subfigure}{0.4\columnwidth}
		\includegraphics[width=\linewidth]{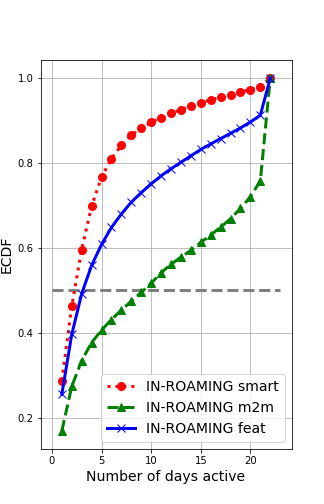}%
		\label{fig:time_active_distribution_inroaming}%
	\end{subfigure}%
	\begin{subfigure}{0.4\columnwidth}
		\includegraphics[width=\linewidth]{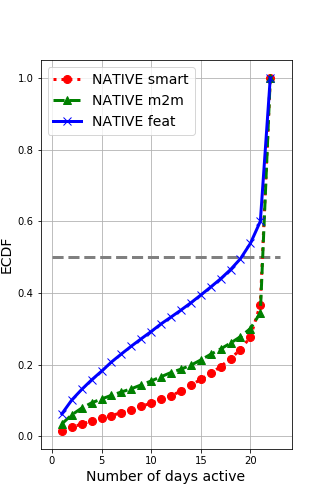}%
		\label{fig:time_active_distribution_native}%
	\end{subfigure}%
	\caption{Number of days devices are active (left) inbound roaming; (right) native. }
	\label{fig:daysActive}
\end{figure}

\subsection{Home Country}

Figure~\ref{fig:homeCountry} shows the distribution of inbound roamers with respect to their home country. 
We first break down the whole population per home country regardless the device class 
(Fig.~\ref{fig:homeCountry}-top).
The top 20 home countries 
contribute more than 93\% of all inbound roaming devices for the MNO, with the top 3 (the Netherlands, Sweden and Spain) 
accounting for about 60\%. 

Figure~\ref{fig:homeCountry}-bottom further detail the breakdown of each home country with respect to the different device classes. 
Columns are ordered to match the histogram in 
Fig.~\ref{fig:homeCountry}-top.
We normalize the values by device class (i.e., per row), using the total number of inbound roaming devices per class, although we show only the top 20 countries, discarding the long tail of the distribution.
We see that 83\% of M2M devices use SIMs from operator from either the Netherlands, Sweden, or Spain; for smarphones and feature phones is 17\% and 35\% respectively. In other words, the distribution per home country for the M2M devices is significantly more skewed than for the other two classes, further corroborating the ``centralization'' of M2M platforms.


\af{...but I don't think they sum to 100, isn't it?}
\ael{no, we only show the top 20 countries, but we normalized using the TOTAL number of m2m devices per class, so they don't sum up to 100}


\subsection{Spatio-Temporal Dynamics}
We further analyze how long M2M devices are active in our dataset. For this, we count the 
overall number of days the device is generating data, voice, or signaling traffic. 
Figure~\ref{fig:daysActive}-left
plots the empirical CDF of the number 
of active days for two device classes, M2M and smartphone devices in the inbound roaming class. 
Considering inbound roamers (left plot), IoT devices (category \emph{"m2m"}) are active 4.5x longer than smartphones as a median (9 days for M2M devices and 2 days for smartphones),
while the 2 device types present similar properties if they are native devices (right plot).
When aggregating this information regardless of the roaming category, we note that M2M devices 
spend less time connected to the network than that smartphones. We conjecture that this can be due to the roaming nature of those devices which enables them 
to switch network if/when needed, or could be the application logic itself which uses the network only when needed.
Unfortunately, the dataset does not offer sufficient details to unravel this aspect.



\begin{figure}[!t] 
	\centering
    \includegraphics[width=0.85\columnwidth]{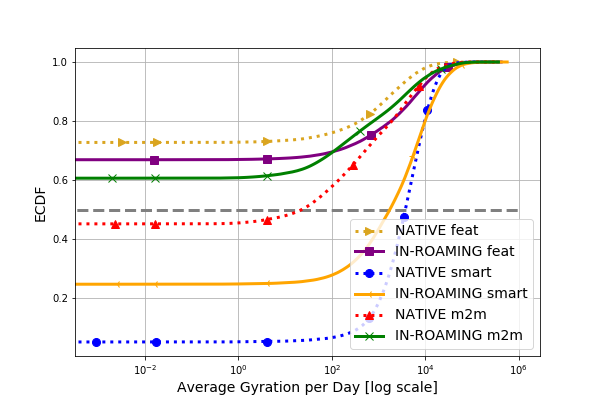}
	\caption{Radius of gyration comparison.}
	\label{fig:mobility}
\end{figure}

In Figure~\ref{fig:mobility} we investigate the mobility of the different device classes.
For this, we evaluate the radius of gyration for the device, capturing the area the device usually travels. 
We use the physical coordinates of the cell sectors to which devices connect to as a proxy of the actual device position, and compute a centroid (an aggregate representation of where in the country the device was located) and the gyration radius (indicating how far from the centroid the device was moving). Both are weighted based on the time spent connected to each cell sector by the devices. We compute daily metrics, and present averages across days.
Results confirm expectation, i.e., the M2M inbound roaming devices are in majority stationary, with only 20\% devices present a gyration larger than 1km (some likely due to cell reselection, rather than actual movements).


%% file: sections/m2m_traffic.tex
\label{sec:m2m_traffic}

In this section, we continue the analysis of the MNO dataset investigating M2M communication patterns. 
We present next how "things" are actually using the cellular network, which is the radio technology on which they depend most and how much traffic they generate.

\begin{figure}[!t]
	\centering
	\begin{subfigure}{0.33\columnwidth}
		\includegraphics[width=\columnwidth]{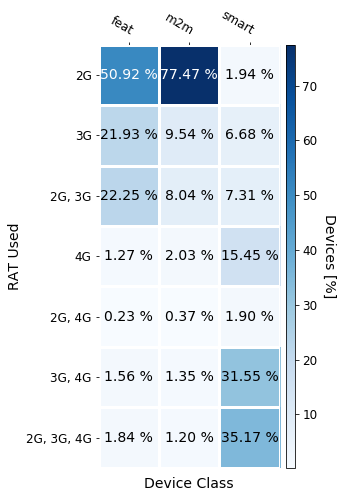}
	     \label{fig:RATusage}%
	\end{subfigure}%
		\begin{subfigure}{0.33\columnwidth}
		\includegraphics[width=\columnwidth]{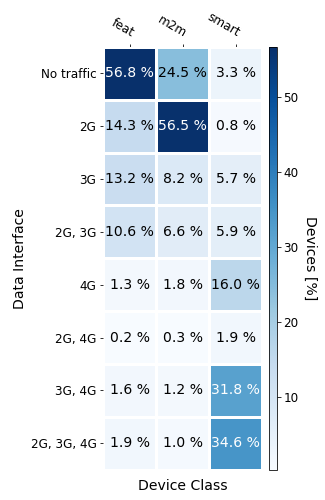}%
		\label{fig:dataUsage}%
	\end{subfigure}%
		\begin{subfigure}{0.33\columnwidth}
		\includegraphics[width=\columnwidth]{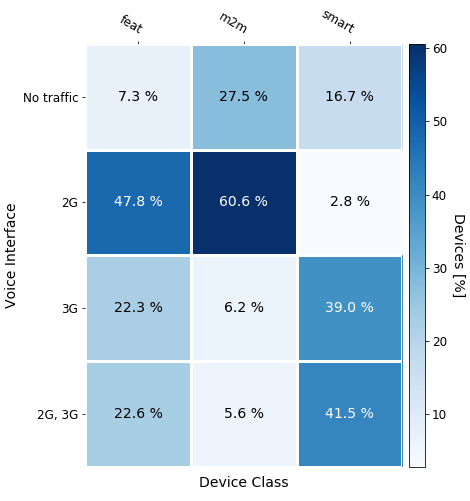}%
		\label{fig:voiceUsage}%
	\end{subfigure}%
	\caption{Devices share with respect to services (left) connectivity; (center) data traffic; (right) voice traffic.}
	\label{fig:network_usage}
\end{figure}

\subsection{Device Network Usage}
Using the device activity on the data or voice interfaces per \ac{RAT}, we generate a view of the patterns for each of the device classes 
(see Fig.~\ref{fig:network_usage}-left).
We find that the vast majority of M2M devices (77.4\%) are active on the 2G network only, while smartphones have mostly 3G and/or 4G capabilities.
Similar to M2M devices, feature phones are also dependent mostly on the 2G radio network (50.9\%). 

\begin{figure*}[!h]
	\centering
	\begin{subfigure}{0.31\textwidth}
		\includegraphics[width=\linewidth]{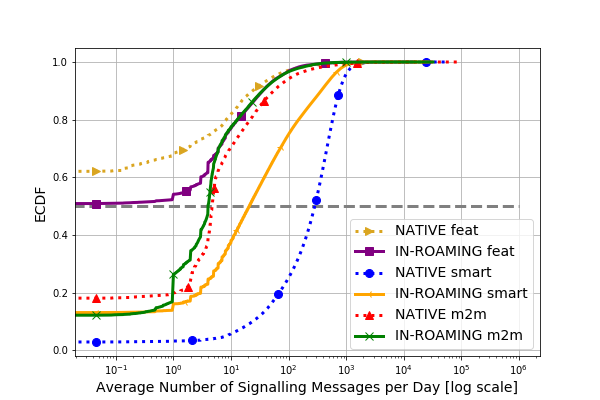}%
		\label{fig:signalling}%
	\end{subfigure}%
		\begin{subfigure}{0.31\textwidth}
		\includegraphics[width=\linewidth]{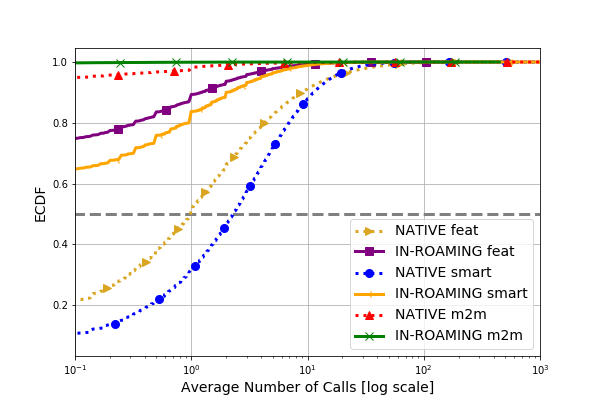}%
		\label{fig:calls}%
	\end{subfigure}%
		\begin{subfigure}{0.31\textwidth}
		\includegraphics[width=\linewidth]{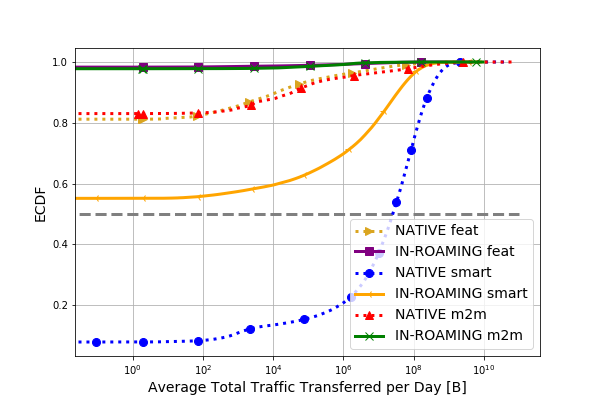}%
		\label{fig:bytes}%
	\end{subfigure}%
	\caption{Traffic analysis for in-roaming and native M2M devices (left) signaling; (center) calls; (right) data usage.} 
	\label{fig:traffic_patterns}
\end{figure*}

When focusing on M2M devices voice usage (Fig.~\ref{fig:network_usage}-right)
we find that 60.6\% use the 2G network, but 27.5\% do not generate any voice traffic. 
Furthermore, when checking just the activity of the devices in the three different classes on the data interfaces 
(Fig.~\ref{fig:network_usage}-center)
of different RATs, we find that 56.7\% of all the M2M devices are indeed only active on the 2G data interface. 
Interestingly, we note that 24.5\% of the M2M devices are actually not active on the data interfaces of the cellular network, relying only on voice communications.\footnote{We use voice services in a broad sense, as M2M devices do not make phone calls, but can use communications similar to SMS.} Notice also how 56.8\% of feature phones do not generate any data traffic, but only 7.3\% of them do not generate voice traffic, i.e., as expected, those type of devices are commonly used for calls.

The sustained dependency of M2M devices and also features phones on the 2G network bring to light the discussion around the need of \acp{MNO} to keep maintaining the legacy technology. Some MNOs (e.g., AT\&T) already shutdown 2G services, while other announced their target dates.\footnote{\url{https://1ot.mobi/blog/2g-and-3g-networks-are-shutting-down-globally}}

\subsection{Traffic Volumes}
We analyse the amount of radio resource management signaling events from devices, number of voice calls and amount of data traffic the M2M devices generate compared to smartphones (Fig.~\ref{fig:traffic_patterns}).
Over the three weeks, we find that the number of resource management events the M2M devices trigger is much smaller than the traffic generated by smartphone devices, regardless their roaming configuration 
(Fig.~\ref{fig:traffic_patterns}-left).
This is partially explained by the fact that M2M devices are also more stationary compared to smartphone devices (Fig.~\ref{fig:mobility}).
Feature phones, however, generate less signaling traffic than even M2M devices as they mostly likely due to the lack of data services usage. 

We further check the average number of voice calls per day for the different device categories (i.e., native M2M, inbound roaming M2M, native smartphones and inbound roaming smartphones). 
In 
Fig.~\ref{fig:traffic_patterns}-center
we show that, although for the vast majority of M2M devices we do not find any calls registered, there is a small fraction for which the number of voice calls is non-null (regardless their roaming configuration). 
We conjecture these might be due to M2M security applications (e.g., emergency elevator services, home security). 

Finally, we analyse the total volume of data traffic  the different categories of devices transferred in different roaming configurations (namely, native and inbound roaming).
Figure~\ref{fig:traffic_patterns}-right
shows that inbound roaming M2M devices generate a very small amount of data traffic, similar to inbound roaming feature phones. 
Some native M2M devices show non-null data traffic usage (20\% of devices generate more than 1 Byte of data in average per day). We note that they have a very similar pattern of data traffic usage with feature phones. 
There is a clear difference in people's behavior while roaming, which we extract from the comparison of smartphones native to the home country of the MNO and the inbound roaming smartphones. 
We assume that the decreased volume of traffic for inbound roamers is due to fear of potential bill shock the users might incur when traveling outside their home country (non-EU).

%
%

%% file: sections/smip.tex
\label{sec:smip}

As traffic dynamics differ between different IoT verticals, in this section we investigate two of the most prominent: smart meters, and connected cars. In particular, we focus on their connectivity and mobility, signaling volume, and data volume.


\subsection{SMIP Device Activity}

We measure SMIP signaling activity looking at the patterns of Attach, Routing Area Update, and Detach signaling procedures that we capture from the passive monitoring of \ac{MSC}  and \ac{MME} elements (see Figure~\ref{fig:mno_architecture}).
Such traffic is known as ``background traffic'', and it
does not bring any profit to the service provider, 
while may lead to possible overheads 
which we show is more significant than for inbound roaming devices.

\begin{figure}[!t]
	\centering
	\begin{subfigure}{0.5\columnwidth}
		\includegraphics[width=\columnwidth]{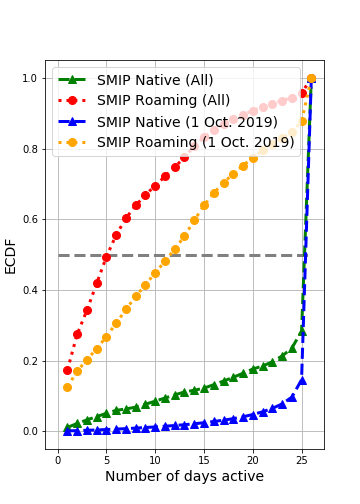}
	     \label{fig:smip_activeDays}%
	\end{subfigure}%
		\begin{subfigure}{0.5\columnwidth}
		\includegraphics[width=\columnwidth]{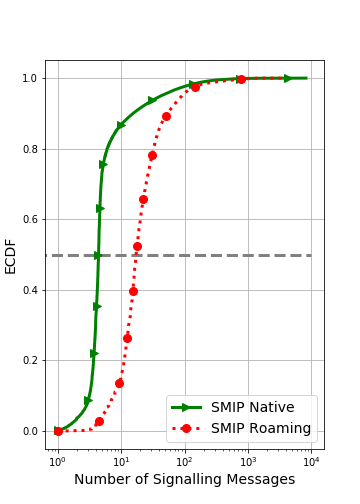}%
		\label{fig:smip_all_signalling}%
	\end{subfigure}%
	\caption{\small Device activity for SMIP Native and SMIP Roaming groups during 1-26 October 2019: a) Total number of days SMIP devices were active during the period we study. We show the active time for total set of devices detected in October 2019, as well as the active time of the devices detected on October 1 across the entire period of analysis.  b) Average number of signaling messages per SMIP device per day. 
	}
	\label{fig:device_activity_rat}
\end{figure}

\begin{figure*}[!h]
	\centering
	\begin{subfigure}{0.31\textwidth}
		\includegraphics[width=\linewidth]{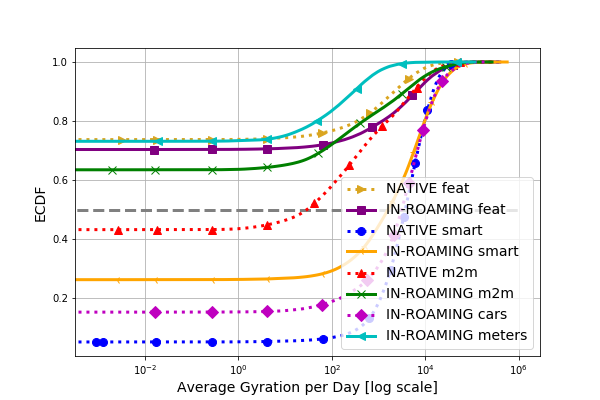}
	     \label{fig:vertical_mobility}%
	\end{subfigure}%
		\begin{subfigure}{0.31\textwidth}
		\includegraphics[width=\linewidth]{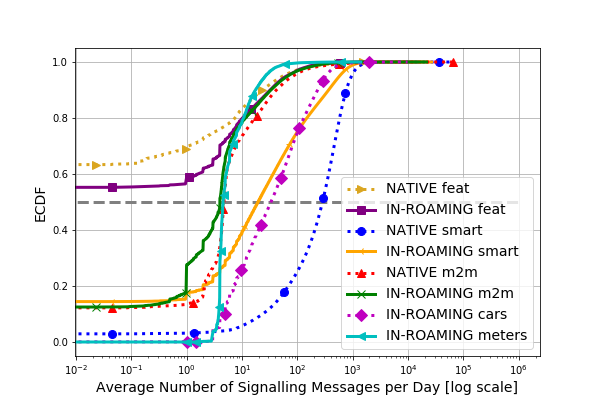}%
		\label{fig:vertical_signaling}%
	\end{subfigure}%
		\begin{subfigure}{0.31\textwidth}
		\includegraphics[width=\linewidth]{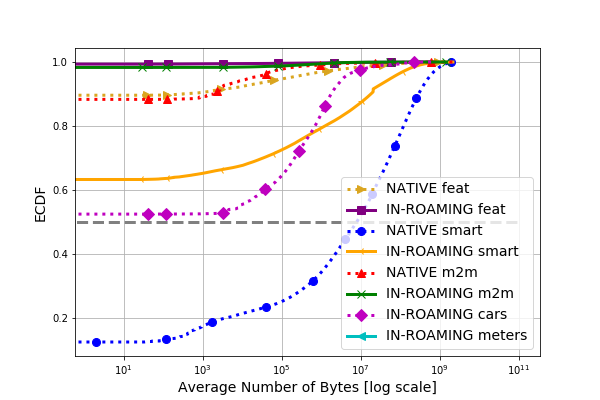}%
		\label{fig:vertical_data}%
	\end{subfigure}%
	\caption{Connected cars and smart meters traffic patterns (left) mobility; (center) signaling; (right) data usage.}
	\label{fig:verticals_usage}
\end{figure*}

Figure~\ref{fig:device_activity_rat}(left) reports on the number of days SMIP devices have been active, i.e., their triggered at least one signaling event per day (either on data or voice interfaces). We split devices between native and inbound roamers, and report on their activity across the whole time period, as well as on the devices being active from the first day of the time period.

We can see that native devices have long-lasting connectivity (73\% are active for the whole period), while the opposite is true for roaming devices (50\% are active only up to 5 days).
We conjecture that this is a side-effect of the fundamentally different manner in which they connect to radio resources: 
roaming devices are free to connect to any UK operator, while native devices rely exclusively on the MNO we study.  
Figure~\ref{fig:device_activity_rat}(left) also shows the effect of the ongoing deployment of SMIP devices. Notice indeed how the fraction of constantly active native devices increase to 83\% when considering the ones active on the first day of the dataset.

 Figure~\ref{fig:device_activity_rat}(right) reports instead on the generated background traffic. Interestingly, notice how roaming SMIP 
generates on average ten times more signaling messages than a native ones.
This considers all the signaling events associated with the smart meters, regardless these procedures being successful or not. 
When considering only the failed events, only 10\% of all SMIP devices registered to the MNO during October 2019 had at least one failed signaling message, but this increase to 35\% when considering roaming devices. Unfortunately, we do not have sufficient data to understand if the increased background traffic is a side effect of roaming, or the roaming itself is a symptom or something deeper, such as network coverage issues.

Looking at the supported radio technologies (see Section~\ref{sec:dataset}), all SMIP roaming devices are only 2G capable; this is confirmed also looking that the RAT used by the devices. Conversely, native SMIP support both 2G and 3G, but 2/3 of them use only on 3G, while the rest uses both 2G and 3G connectivity. 

\subsection{Traffic Analysis for IoT Verticals}

Using the exposed \ac{APN} information from inbound roaming IoT devices in our MNO dataset, we separate devices mapping to connected cars. We further use this dataset to contrast against the traffic patterns of smart energy meters. 
In Figure~\ref{fig:verticals_usage} our analysis shows that connected cars are very similar to normal inbound roaming smartphones, with high mobility patterns 
(left) large volume of signaling traffic 
(center) and data traffic 
(right).
At the same time, smart energy meters are stationary IoT devices demonstrate a completely different behavior. 
As expected, they are stationary devices that generate very little signaling traffic as well as data traffic, when compared to the connected cars. 
These patterns validate our intuition on the manner in which these two groups of IoT devices use the visited network.

%% file: sections/discussion.tex
\label{sec:discussion}

Our analysis focuses on the role of international roaming to support the growing IoT. 
We combined two datasets, one from an operational M2M platform covering multiple countries, and the other from an operational MNO,
to shed light on the dynamics around roaming for M2M communications. 
This is, to the best of our knowledge, the first analysis of how roaming supports IoT/M2M connectivity world-wide, complementing prior work, which brought a limited view of roaming or overlooked it altogether. 
Despite the novelty of our analysis, our view is still limited to the footprint of the system we analyze. 
Thus, our analysis has a strong European focus, 
where roaming is heavily used and regulation efforts are pursuing the awakening of ``silent roamers''.\footnote{https://www.evolved-intelligence.com/products/roaming/silent}

Although different technical roaming configurations (i.e., \ac{HR}, 
\ac{LBO}, \ac{IHBO}) might be used for different IoT verticals (allowing the 
M2M platform to respond to specific QoS requirements), from the M2M platform dataset we currently lack visibility into these details. 
We complement this analysis 
investigating the traffic of more than 3 million UK smart meters
comparing the ones configured in \ac{HR} roaming with ones native to the MNO.
This IoT vertical account for the largest number of devices compared to the other IoT verticals we were able to identify (e.g., connected cars). 
Previous work characterized different verticals such wearables~\cite{kolamunna2018first} or connected 
cars~\cite{andrade2017connected} highlighted the difference in terms of 
requirements of these applications and the corresponding devices used, but did not highlight their reliance on roaming.  

Finally, our analysis of the global M2M platform relies on 4G signaling information we collected from 120,000 IoT devices. 
As MNOs across the world move to phase out 2G/3G support, IoT verticals will likely rely on more sustainable 
technologies such as 4G/LTE, driven in part by sectors like the connected automotive industry, 
in which seamless, cross-border, ultra-reliable, low-latency connectivity is of paramount importance.
In countries such as Japan, South Korea, Singapore or Australia, MNOs have already switched off 2G.
MNOs in Europe are reportedly planning to retire their legacy 2G/3G networks starting 2020.  
However, our results from characterizing the IoT devices connected to the European MNO 
show that IoT devices such as smart meters are currently active mostly in 2G or 3G networks.
Thus, while informative, we infer that the global footprint of the M2M platform we 
provide is a lower-bound.

Given its power to support the commercial success of IoT verticals, roaming is coming 
to other IoT technologies as well. For example, \ac{NB-IoT} is a low-power wide-area 
network technology developed for the huge volume and concentration of connected ``things'' 
that receive and transmit only small amounts of data, but do so over long periods of time, such as smart meters. 
The GSMA announced the first international NB-IoT roaming trial back in June 2018, with 
numerous others having taken place since. The planned deployment of NB-IoT coupled with 
roaming support will likely create a powerful environment to support additional growth of 
IoT, as operators, device manufacturers and vertical sectors seek a means of delivering a 
low-cost, future-proof IoT. NB-IoT will enable visited MNOs to easily detect the inbound 
roaming IoT devices, a task that currently is challenging.

%% file: sections/conclusions.tex
\label{sec:conclusions}


In this paper, we exposed the role international carriers have in enabling IoT verticals, and offered the first characterization of a global M2M platform.
We also showed how such solutions leverage the maturity of the roaming infrastructure to provide the reliability 
and ubiquity to sustain consistent communications for IoT verticals such as smart energy meters. 
For example, we revealed that using IoT SIMs, the M2M platform supports 
IoT verticals in over 70 countries over 4G networks. 

Despite its exponential growth, IoT also translates into increased stress on the 
infrastructure of the MNOs to which they connect. Our analysis of the device 
population of a European MNO showed that M2M devices account for 26\% (10.1M) of all connected devices
across 3 weeks, out of which 75\% are inbound roamers.
We uncovered that the inbound roaming IoT devices we identify are connected for longer 
periods than people roaming into the same network. At the same time, these devices 
generate very little traffic. In other words, though these devices occupy radio 
resources in the MNO's network and exploit the MNO's interconnections in the cellular 
ecosystem, they do not generate traffic that allows the \ac{MNO} to retrieve the 
corresponding revenue. We further noted that these devices also put stress on the 
MNO a part of the international roaming ecosystem (i.e., MNO interconnection signaling 
through a roaming hub, data and financial clearing). In a market expected to reach 75.44 billion worldwide by 2025, i.e., almost 10x the estimated world population, this puts in perspective the importance of the M2M platform and the corresponding international carrier in 
supporting the relationships between \ac{VMNO}s and \ac{IoT} verticals.